\documentclass[a4paper,11pt]{amsart}
\usepackage{graphicx}
\usepackage{amssymb}
\begin{document}

\hyphenation{gra-vi-ta-tio-nal re-la-ti-vi-ty Gaus-sian
re-fe-ren-ce re-la-ti-ve gra-vi-ta-tion Schwarz-schild
ac-cor-dingly gra-vi-ta-tio-nal-ly re-la-ti-vi-stic pro-du-cing
de-ri-va-ti-ve ge-ne-ral ex-pli-citly des-cri-bed ma-the-ma-ti-cal
de-si-gnan-do-si coe-ren-za pro-blem gra-vi-ta-ting geo-de-sic
per-ga-mon cos-mo-lo-gi-cal gra-vity cor-res-pon-ding
de-fi-ni-tion phy-si-ka-li-schen ma-the-ma-ti-sches ge-ra-de
Sze-keres con-si-de-red tra-vel-ling ma-ni-fold re-fe-ren-ces
geo-me-tri-cal in-su-pe-rable sup-po-sedly at-tri-bu-table
Bild-raum in-fi-ni-tely counter-ba-lan-ces iso-tro-pi-cally
pseudo-Rieman-nian cha-rac-te-ristic geo-de-sics
Koordinaten-sy-stems}

\title[On the gravitational redshifts of spectral lines] {{\bf On the gravitational redshifts\\of
spectral lines\\\emph{A critico-historical study}}}

\author[Angelo Loinger]{Angelo Loinger}
\address{A.L. -- Dipartimento di Fisica, Universit\`a di Milano, Via
Celoria, 16 - 20133 Milano (Italy)}
\author[Tiziana Marsico]{Tiziana Marsico}
\address{T.M. -- Liceo Classico ``G. Berchet'', Via della Commenda, 26 - 20122 Milano (Italy)}
\email{angelo.loinger@mi.infn.it} \email{martiz64@libero.it}

\vskip0.50cm

\begin{abstract}
A beautiful, detailed computation by Whittaker has enabled us to
prove in a rigorous way that the gravitationally redshifted
frequency of a monochromatic e.m. wave sent forth at the surface
of a celestial body is propagated \emph{unaltered} from the
emitting source to terrestrial observers. We remark that in the
customary treatments only qualitative and inaccurate
justifications of this fact are given.
\end{abstract}

\maketitle

\vskip1.20cm \noindent \small \textbf{Summary} -- \textbf{1} Aim
of the paper. -- \textbf{2}. Observational and experimental data.
-- \textbf{3}. The usual deduction of the gravitational redshift,
and its weak point. -- \textbf{4}. Electromagnetic equations in an
arbitrary gravitational field. -- \textbf{5}. Electromagnetic
equations in the field of a gravitating body. -- \textbf{6}.
Spherical e.m. waves starting from a gravitating body. --
\textbf{7}. Waves of very high frequency. -- \textbf{7bis}. The
eikonal equation as the equation of the characteristics of
Maxwell's wave equations. -- \textbf{8}. Waves of arbitrary
frequency. -- \emph{Appendix}: Geometrical optics in GR --
Maxwellian and Einsteinian characteristic surfaces. --

\normalsize

\vskip1.20cm \noindent \textbf{1.} -- As it is well known, there
are three kinds of redshift: purely gravitational, cosmological,
generated by a Doppler effect. In this paper we shall clarify a
subtle point of the usual treatments of the first instance.

\vskip1.20cm \noindent \textbf{2.} -- For the observational data
on the astrophysical gravitational redshifts, see the recent
article by Pasquini \emph{et alii} \cite{1}. We limit ourselves to
emphasize that evident gravitational redshifts have been detected
in spectral lines of X-ray bursts on neutron-star surfaces
\cite{2}.

\par Interesting terrestrial experiments, based on the Principle
of Equivalence, have been made by Pound \emph{et alii} in 1960 and
in 1964 \cite{3}.

\vskip1.20cm \noindent \textbf{3.} -- For the theoretical
definition of the gravitational redshift, it is suitable to
consider \emph{static} gravitational potentials $g_{jk}$,
$(j,k=0,1,2,3)$, referred to \emph{static} coordinate systems.

\par Let $\nu$ be the frequency of a monochromatic radiation sent
forth by a given atom (at rest) at some point $P$ on the surface
of the celestial body, which generates the field $g_{jk}$. We
have:

\begin{equation} \label{eq:one}
\frac{1}{c} \, \textrm{d}s = \sqrt{g_{00}(P)} \, \textrm{d}t_{P}
 \quad;
\end{equation}

for the radiation of frequency $\nu'$, emitted by an identical
atom at a point $P'$ on the Earth, we can write analogously:

\begin{equation} \label{eq:oneprime}
\frac{1}{c} \, \textrm{d}s = \sqrt{g_{00}(P')} \, \textrm{d}t_{P'}
\tag{1$'$} \quad,
\end{equation}

from which:

\begin{equation} \label{eq:two}
\nu = \nu' \frac{\sqrt{g_{00}(P)}}{\sqrt{g_{00}(P')}} \quad,
\end{equation}

since $\nu / \nu'=\textrm{d}t_{P} / \textrm{d}t_{P'}$. We see that
$\nu<\nu'$, because $g_{00}(P)<g_{00}(P')$: gravitational
redshift.

\par We recall that the wavelength $\lambda$ and the frequency
$\nu$ satisfy the relation $\lambda\nu=V(t)$, if $V(t)$ is the
propagation velocity of the e.m. wave through the
pseudo-Riemannian spacetime. At the arrival on Earth, $V(t)
\approx c$.

\par In a Newtonian approximation, we have $g_{00}=1-2U/c^{2}$,
where $U$ is the Newtonian potential which satisfies the equation
$\nabla^{2}U=-4 \pi G \mu$, if $\mu$ is the mass density.
Accordingly, $\sqrt{g_{00}} \approx 1-U/c^{2}$, and we get:

\begin{equation} \label{eq:twoprime}
\frac{\nu}{\nu'} = \frac{1-U(P)/c^{2}}{1-U(P')/c^{2}}  \tag{2$'$}
\quad;
\end{equation}

from which:

\begin{equation} \label{eq:twosecond}
\frac{\nu-\nu'}{\nu'} \approx \frac{U(P')}{c^{2}} -
\frac{U(P)}{c^{2}} \tag{2$''$}\quad.
\end{equation}

It is customary to assert -- as a consequence of generic, or
seemingly intuitive, arguments -- that the frequency $\nu$ of the
considered radiation arrives \emph{unaltered} at the terrestrial
observers. Only Weyl \cite{4} and v. Laue \cite{5} remark, but
without an explicit proof, that \emph{by virtue of Maxwell
equations} the e.m. monochromatic waves are propagated in a
pseudo-Riemannian spacetime without any modification of their
frequencies.

\par By making use of a result by Whittaker \cite{6}, we shall
demonstrate with an accurate computation that effectively the
frequency $\nu$ of the considered wave traverses unchanged the
distance between the celestial body and the Earth.

\vskip1.20cm \noindent \textbf{4.} -- With Whittaker \cite{6}, we
consider the ideal case in which the gravitational action of the
e.m. field $F_{jk}$, $(j,k=0,1,2,3)$, -- action which is in
general small -- can be ignored. If $\Phi_{j}$ is the e.m.
4-potential, we have:

\begin{equation} \label{eq:three}
F_{jk} = \Phi_{j:k} - \Phi_{k:j} = \Phi_{j,k} - \Phi_{k,j} \quad,
\end{equation}

where a colon and a comma denote respectively a covariant and an
ordinary derivative. If $j_{k}$ is the 4-current, Maxwell
equations tell us that

\begin{equation} \label{eq:four}
g^{jk} \, \Phi_{n:j:k} + R_{n}^{j} \, \Phi_{j} = j_{n} \quad ;
\quad (n=0,1,2,3) \quad ,
\end{equation}

if $R_{jk}$ is the contracted curvature tensor. The deduction of
eqs. (\ref{eq:four}) implies the use of covariant Lorentz
condition $g^{jk}\Phi_{j:k}=0$.

\par We shall consider the gravitational filed about a
gravitating centre with a given mass $M$, and e.m. fields acting
in the spatial region in which Ricci tensor $R_{jk}$ is equal to
zero. Accordingly, eqs. (\ref{eq:four}) become:

\begin{equation} \label{eq:five}
\Box \, \Phi_{n} = j_{n} \quad ; \quad (n=0,1,2,3) \quad ,
\end{equation}

if $\Box \equiv g^{jk}(\ldots)_{:j:k}$.

\par The equations for $F_{jk}$, which correspond to eqs.
(\ref{eq:four}) for $\Phi_{n}$, are \cite{7}:

\begin{equation} \label{eq:fourprime}
\Box \, F_{pq} + R_{p}^{m} F_{mq} - R_{q}^{m} F_{mp} -2
R_{pqlm}\,F^{ml} = j_{p:q} - j_{q:p} \tag{4$'$} \quad ;
\end{equation}

and for the region in which $R_{jk}=0$:

\begin{equation} \label{eq:fiveprime}
\Box \, F_{pq} -2 R_{pqlm}\,F^{ml} = j_{p:q} - j_{q:p} \tag{5$'$}
\quad .
\end{equation}

\vskip1.20cm \noindent \textbf{5.} -- Whittaker \cite{6} starts
from the metric of the Schwarzschild manifold as described by the
standard (Hilbert-Droste-Weyl) form of the interval $\textrm{d}s$
(where $\alpha \equiv 2GM/c^{2}$):

\begin{equation} \label{eq:six}
\textrm{d}s^{2} =
\left(\frac{r-\alpha}{r}\right)c^{2}\textrm{d}t^{2} -
\left(\frac{r}{r-\alpha}\right)\textrm{d}r^{2} - r^{2}\left(
\textrm{d}\vartheta^{2} + \sin^{2}\vartheta \,
\textrm{d}\varphi^{2}\right) \quad .
\end{equation}

He calculates the covariant derivatives $\Phi_{n:j}$ and
$\Phi_{n:j:k}$; the substitution of their values in eqs.
(\ref{eq:five}) gives a system of four partial differential
equations for determining $\Phi_{0}$, $\Phi_{1}$, $\Phi_{2}$,
$\Phi_{3}$. (see eqs. (21), (22), (23), (24) of \cite{6}).

\vskip1.20cm \noindent \textbf{6.} -- We consider systems of e.m.
spherical waves with the centre at the centre of the gravitating
spherical body, and having origin (or ending up) at the surface
$(r>\alpha$) of this body. With Whittaker, we specialize his eqs.
(21)--(22)--(23)--(24) for the instance
$\Phi_{0}=\Phi_{1}=\Phi_{2}=0$, $\Phi_{3} \neq 0$. Also the
4-current $j_{n}$ is zero, except at the origin of the waves. In
this way, only eq. (24) of \cite{6} survives, in the following
simpler form:

\begin{equation} \label{eq:seven}
\frac{r}{c^{2}(r-\alpha)} \, \frac{\partial^{2} \Phi_{3}}{\partial
t^{2}} - \frac{r-\alpha}{r} \, \frac{\partial^{2}
\Phi_{3}}{\partial r^{2}} - \frac{\alpha}{r^{2}} \frac{\partial
\Phi_{3}}{\partial r}  - \frac{1}{r^{2}} \frac{\partial^{2}
\Phi_{3}}{\partial \vartheta^{2}} + \frac{1}{r^{2}} \, \cot
\vartheta \, \frac{\partial \Phi_{3}}{\partial \vartheta} = 0
\quad .
\end{equation}

We require a $\Phi_{3}$ of the following kind (Whittaker denotes
with $p$ the pulsation $\omega=2\pi\nu$):

$\exp(i \omega t) \times (\textrm{a function of } r \textrm{
only}) \times (\textrm{a function of } \vartheta \textrm{ only})$
-- say,

\begin{equation}\label{eq:eigth}
\Phi_{3} = \exp(i \omega t) \, f(r) \, g(\vartheta) \quad .
\end{equation}

By substituting (\ref{eq:eigth}) in (\ref{eq:seven}), we obtain:

\begin{equation} \label{eq:nine}
\frac{-\omega^{2}r}{c^{2}(r-\alpha)} -
\frac{r-\alpha}{r}\frac{1}{f(r)}
\frac{\textrm{d}^{2}f(r)}{\textrm{d}r^{2}} -
\frac{\alpha}{r^{2}}\frac{1}{f(r)}\frac{\textrm{d}f(r)}{\textrm{d}r}-
\frac{1}{r^{2}}\frac{1}{g(\vartheta)}\frac{\textrm{d}^{2}g(\vartheta)}{\textrm{d}\vartheta^{2}}+
\frac{1}{r^{2}}\frac{\cot\vartheta}{g(\vartheta)}\frac{\textrm{d}g(\vartheta)}{\textrm{d}\vartheta}
\, .
\end{equation}

With a classic procedure, Whittaker finds that

\begin{equation} \label{eq:ten}
g(\vartheta) = \sin^{2}\vartheta \,
\frac{\textrm{d}P_{n}(\cos\vartheta)}{\textrm{d}(\cos\vartheta)}
\quad ,
\end{equation}

and $f(r)$ satisfies the equation

\begin{equation} \label{eq:eleven}
\frac{r-\alpha}{r} \, \frac{\textrm{d}^{2}f}{\textrm{d}r^{2}} +
\frac{\alpha}{r^{2}} \frac{\textrm{d}f}{\textrm{d}r} + \left\{
\frac{\omega^{2}r}{c^{2}(r-\alpha)} - \frac{n(n+1)}{r^{2}}\right\}
f = 0 \quad
\end{equation}

where $n=0,1,2,\ldots$; eq. (\ref{eq:eleven}) is similar to
Mathieu's equation \cite{8}, its solution $f(r)$ will be
calculated in sects. \textbf{7}. and \textbf{8}.

\vskip1.20cm \noindent \textbf{7.} -- When the e.m. waves are of a
very high frequency, so that the pulsation $\omega$ is very large,
we can neglect in eq. (\ref{eq:eleven}) the term $-n(n+1)/r^{2}$
in comparison with $\omega^{2}r/c^{2}(r-\alpha)$, and thus eq.
(\ref{eq:eleven}) becomes

\begin{equation} \label{eq:twelve}
\frac{r-\alpha}{r} \, \frac{\textrm{d}^{2}f}{\textrm{d}r^{2}} +
\frac{\alpha}{r^{2}} \frac{\textrm{d}f}{\textrm{d}r} +
\frac{\omega^{2}r}{c^{2}(r-\alpha)} \, f = 0 \quad ;
\end{equation}

the solution is:

\begin{equation} \label{eq:thirteen}
f = A \exp (i\omega r/c) \cdot (r-\alpha)^{i\omega\alpha\ /c} + B
\exp (-i\omega r/c) \cdot (r-\alpha)^{-i\omega\alpha /c} \quad .
\end{equation}

with $A$, $B$ arbitrary constants. Therefore we can write:

\begin{equation} \label{eq:fourteen}
\Phi_{3} = \exp (i\omega t) \cdot \exp (\pm i\omega r/c)
(r-\alpha)^{\pm i\omega\alpha /c} \cdot \sin^{2}\vartheta \,
\frac{\textrm{d}P_{n}(\cos\vartheta)}{\textrm{d}(\cos\vartheta)}
\quad ;
\end{equation}

the signs plus and minus denote, respectively, convergent or
divergent waves. If we write eq. (\ref{eq:fourteen}) in the form

\begin{equation} \label{eq:fourteenprime}
\Phi_{3} = [\textrm{a function of } (t\pm h(r)] \times [\textrm{a
function of } \vartheta ] \tag{14$'$} \quad ,
\end{equation}

we see that the wave velocity at the point $(r,\vartheta)$ is
$1/(\textrm{d}h/\textrm{d}r)$; now, eq. (\ref{eq:fourteen}) tells
us that $h(r)=(r/c)+(\alpha /c) \ln |r-\alpha|$, from which
$\textrm{d}h / \textrm{d}r=r/c(r-\alpha)$; accordingly:

\begin{equation} \label{eq:fifteen}
\left|\frac{\textrm{d}r}{\textrm{d}t} \right| = c \left(
1-\frac{\alpha}{r}\right) \quad .
\end{equation}

This value coincides with the velocity of the \emph{light-rays} as
null geodesics of metric (\ref{eq:six}); indeed,
$\textrm{d}s^{2}=0$ gives:

\begin{equation} \label{eq:sixteen}
\frac{r-\alpha}{r} \, \textrm{d}t^{2} - \frac{r}{c^{2}(r-\alpha)}
\, \textrm{d}r^{2}=0 \quad ,
\end{equation}

from which eq. (\ref{eq:fifteen}). As we shall see in the sequel,
this coincidence is not casual.

\vskip1.20cm \noindent \textbf{7bis.} -- We recall that, both in
the pre-relativistic and in the relativistic physics, the notion
of geometrical optics is susceptible of two different
interpretations. First interpretation: the eikonal equation
represents the approximation of the wave equation for a very high
frequency; the general-relativistic proof of this result is due to
v. Laue \cite{9}. The procedure of previous section \textbf{7}
applies essentially this viewpoint. Second interpretation: the
eikonal equation gives the propagation law of the wave-fronts,
without any reference to wave frequencies; the
general-relativistic proof of this result is due to Whittaker
\cite{10}.

\par He started from eqs. (\ref{eq:four}), written in the
following form:

\begin{equation} \label{eq:seventeen}
g^{jk} \, \frac{\partial^{2} \Phi_{n}}{\partial x^{j} \partial
x^{k}}+ (\textrm{terms not involving second derivatives of the }
\Phi_{n} \textrm{'s}) = 0 \quad.
\end{equation}

His purpose was to find the characteristic surfaces (that are
discontinuity -- or singular -- surfaces) of eqs. (\ref{eq:four}).
Now, formula (\ref{eq:seventeen}) recalls that the characteristics
of a partial differential equation of the second order depend only
on the terms involving the second derivatives of the solutions.
Eqs. (\ref{eq:four}) are of the hyperbolic kind, and therefore the
differential equation of the characteristics (see eq.
(\ref{eq:eighteen}), \emph{infra}) gives also the law of motion of
the wave-fronts; it coincides with the eikonal equation of the
first interpretation.

\par The theory of the characteristics tells us that the functions
$z(x^{0}, x^{1}, x^{2}, x^{3})$ describing the characteristic
surfaces $z=0$  of (\ref{eq:seventeen}) are the solutions of the
following differential equation of the first order:

\begin{equation} \label{eq:eighteen}
g^{jk} \frac{\partial z}{\partial x^{j}} \,  \frac{\partial
z}{\partial x^{k}} = 0 \quad .
\end{equation}

The characteristics of (\ref{eq:eighteen}), or bicharacteristics
of (\ref{eq:seventeen}), are the light-rays, \emph{i.e.} the null
geodesics of the manifold, whose interval is given by
$\textrm{d}s^{2}=g_{jk} \, \textrm{d}x^{j} \, \textrm{d}x^{k}$.
Thus, \emph{starting from Maxwell electrodynamics}, Whittaker
found again two fundamental properties of the Einsteinian
gravitation theory.

\par This result has a great conceptual meaning, and an important
consequence, as it was emphasized by Levi-Civita \cite{11} -- see
the \emph{Appendix}.

\par Let us observe that in the first interpretation the
light-rays are represented by very narrow beams of light of very
high frequency -- \emph{i.e.}, by \emph{physical} objects --,
while in the second interpretation they are represented by
\emph{geometrical} trajectories, without any limitation to the
frequency values. Both e.m. interpretations \emph{and} general
relativity give consistently the value (\ref{eq:fifteen}) for the
velocity of the light-rays; the wavelength $\lambda$ does not
remain unchanged as the frequency $\nu$, but is proportional to
velocity. The light-rays are \emph{repulsed} by the gravitating
centre; $|\textrm{d}r/\textrm{d}t|=0$ for $r=\alpha$, and is equal
to $c$ for $r=\infty$.

\par A last remark. It is very easy to see that the
characteristics of eqs. (\ref{eq:fifteen}) coincide with the
characteristics of eqs. (\ref{eq:four}), and are solutions of eq.
(\ref{eq:eighteen}).

\vskip1.20cm \noindent \textbf{8.} -- For the solution $f(r)$ of
eq. (\ref{eq:eleven}) when the frequency of the e.m. waves is
\emph{arbitrary}, Whittaker -- see \S $11$ of \cite{6} -- starts
from the following series:

\begin{equation} \label{eq:nineteen}
f(r) = \exp (\pm \, i\omega r/c) \cdot (r-\alpha)^{\pm \,
i\omega\alpha /c} \left\{ 1+\frac{h_{1}(r)}{\omega}+
\frac{h_{2}(r)}{\omega^{2}}+ \frac{h_{3}(r)}{\omega^{3}}+ \ldots
\right\}\, ;
\end{equation}

then, with some passages, he obtains from eq. (\ref{eq:eleven})
that

\begin{equation} \label{eq:twenty}
h_{s+1}(r) = \frac{1}{2} \, ic \left\{
\Big(1+\frac{\alpha}{r}\Big) \,
\frac{\textrm{d}h_{s}(r)}{\textrm{d}r} +n(n-1)\right\}
\int_{r}^{\infty} \frac{h_{s}(r)\textrm{d}r}{r^{2}}\quad ,
\end{equation}

from which:

\begin{equation} \label{eq:twentyone}
h_{1}(r) = \frac{n(n+1)\,ic}{2r} \, ; \quad h_{2}(r) = -
\frac{n(n+1)\,c^{2}}{4} \left\{ \frac{(n+2)(n-1)}{2r^{2}} +
\frac{\alpha}{r^{3}}\right\} \, , \textrm{\emph{etc}.} \, ;
\end{equation}

with this determination of the series of eq. (\ref{eq:nineteen}),
the e.m. potential $\Phi_{3}$ of eq. (\ref{eq:eigth}) is
completely known:

\begin{equation} \label{eq:twentytwo}
\Phi_{3}(r,\vartheta) =  \exp (i\omega t) \, f(r) \,
\sin^{2}\vartheta \cdot
\frac{\textrm{d}P_{n}(\cos\vartheta)}{\textrm{d}(\cos\vartheta)}
\quad .
\end{equation}

We see that also the e.m. waves of \emph{arbitrary} frequencies
are propagated in a pseudo-Riemaniann manifold with
\emph{unaltered} frequencies, and with the velocity given by eq.
(\ref{eq:fifteen}).

\par It is interesting to remark that a computation concerning
spherical e.m. waves \emph{only} dependent on the radial
coordinate $r$, would give immediately the above result, since the
solution $f(r)$ of

\begin{equation} \label{eq:twentythree}
\frac{r-\alpha}{r} \, \frac{\textrm{d}^{2}f}{\textrm{d}r^{2}} +
\frac{1}{r^{2}} \, \frac{\textrm{d}f}{\textrm{d}r} +
\frac{\omega^{2}\,r}{c^{2}(r-\alpha)} \, f = 0\quad ,
\end{equation}

is simply the $f(r)$ of eq. (\ref{eq:thirteen}).

\newpage
\begin{center}
\noindent \small \emph{\textbf{APPENDIX}}
\end{center} \normalsize

\vskip0.40cm \noindent We give here a r\'esum\'e of the formalism
and of the main properties that characterize the geometrical
optics in GR.

\vskip0.80cm \noindent \textbf{\S1.} -- Let us consider the
spacetime interval $\textrm{d}s$ in a generic pseudo-Riemannian
manifold:

\begin{equation} \label{eq:A1}
\textrm{d}s^{2} = g_{jk}(x^{0},x^{1},x^{2},x^{3}) \,
\textrm{d}x^{j}\,\textrm{d}x^{k} \tag{A.1} \quad , \quad
(j,k=0,1,2,3) \quad ,
\end{equation}

where $g_{00}>0$, and
$g_{\alpha\beta}\,\textrm{d}x^{\alpha}\,\textrm{d}x^{\beta}$,
$(\alpha,\beta=0,1,2,3)$, is negative definite.

\par Two basic equations ($\sigma$ is an affine parameter, and
$x\equiv(x^{0},x^{1},x^{2},x^{3})$):

\begin{equation} \label{eq:A2}
\quad \quad \quad L:= g_{jk} \,
\frac{\textrm{d}x^{j}}{\textrm{d}\sigma} \,
\frac{\textrm{d}x^{k}}{\textrm{d}\sigma} = 0 \tag{A.2} \quad ,
\quad \textrm{(Lagrange-Monge)} \quad ;
\end{equation}

\begin{equation} \label{eq:A3}
\, H:= \frac{1}{2} \, g^{jk}\, \frac{\partial z(x)}{\partial
x^{j}} \, \frac{\partial z(x)}{\partial x^{k}} = 0 \tag{A.3} \quad
, \quad \textrm{(Hamilton-Jacobi)} \quad ;
\end{equation}

Lagrange equations $(\dot{x}^{j}:=\textrm{d}x^{j} /
\textrm{d}\sigma)$

\begin{equation}
\frac{\partial L}{\partial x^{j}} - \frac{\partial}{\partial
\sigma} \left(\frac{\partial L}{\partial \dot{x}^{j}}\right) = 0
\label{eq:A4} \tag{A.4}
\end{equation}

give the \emph{characteristic lines} of (\ref{eq:A2}), which
coincide with the \emph{null} geodesics.

\par Hamilton equations $(p_{j}:=\partial z(x) / \partial x^{j})$

\begin{equation} \label{eq:A5} \tag{A.5}
\dot{x}^{j} = \frac{\partial H}{\partial p_{j}} \quad ; \quad
\dot{p}^{j} = -\frac{\partial H}{\partial x^{j}}
\end{equation}

give the \emph{characteristic lines of} (\ref{eq:A3}), which
coincide with the \emph{null} geodesics.

\par There si a \emph{unique} physical interpretation of these
results, that holds both for $g_{jk}$'s dependent on, and for
$g_{jk}$'s independent of, the time coordinate $x^{0}$: the null
geodesics represent the \emph{light-rays}, equation $z(x)=0$
represents the wave-front of an \emph{electromagnetic} wave.
Remark that \emph{no} use of Einstein field equations has been
made.

\vskip0.80cm \noindent \textbf{\S2.} -- In 1930 Levi-Civita
\cite{11} discovered that the differential equation of the
\emph{characteristic surfaces} of Einstein field equations
coincides with eq. (\ref{eq:A3}). He concluded immediately that a
metric tensor with an undulatory nature is simply the
\emph{support} of an \emph{e.m.} disturbance -- in accord with the
fact that the metric tensor ``is'' the spacetime. This conclusion
was corroborated by Whittaker's proof \cite{10} that eq.
(\ref{eq:A3}) is also the differential equation of the
characteristic surfaces of Maxwell's wave equations. On the other
hand, from the mathematical standpoint the equality

\begin{equation}
\label{eq:A6} g^{jk} \left[z(x)\right] \, \frac{\partial
z(x)}{\partial x^{j}} \, \frac{\partial z(x)}{\partial x^{k}} = 0
\tag{A.6}
\end{equation}

is not an equation, but a trivial identity.

\par Levi-Civita's interpretation is reinforced by the following
facts: \

\emph{i}) No motion of masses generates undulatory $g_{jk}$'s
\cite{12}; \

\emph{ii}) For $G\rightarrow 0$, eq. (\ref{eq:A1}) gives
Minkowski's interval, which admits only an e.m. interpretation; \

\emph{iii}) The absence of a set of privileged reference systems
in GR tells us that the wave nature of a given $g_{jk}$ can be
always destroyed by a suitable change of spacetime coordinates; \

\emph{iv}) Last but not least: as Levi-Civita observed, an
undulatory $g_{jk}$ , which is solution of Einstein equations
$R_{jk}=0$, does not have a true energy-tensor. --

\par An inappropriate analogical comparison with Maxwell
electrodynamics has generated the conviction that undulating
$g_{jk}$'s have a \emph{physical} reality.

\vskip0.80cm \noindent \textbf{\S3.} -- Let us write Einstein
equations in a reference system of \emph{harmonic} coordinates
$y^{0}, y^{1}, y^{2}, y^{3}$; we get:

\begin{equation}\label{eq:A7}
\frac{1}{2} \, g^{mn} \, \frac{\partial^{2}g^{jk}}{\partial
y^{m}\,
\partial y^{n}} - g^{mr} g^{ns} \, \Gamma_{rs}^{j} \, \Gamma_{mn}^{k} =
 \kappa \left( T^{jk} - \frac{1}{2}\, g^{jk} \, T \right) \tag{A.7} \quad ;
\end{equation}

the general theory of the characteristics affirms that the
differential equation of the characteristic surfaces depends only
on the terms containing the derivatives of the highest order -- in
our case, the d'Alembertian terms of eqs. (\ref{eq:A7}), from
which eq. (\ref{eq:A3}).

\par Now, we could obtain this \emph{same} equation for any field, say
$U^{jk\ldots}$, whose equations of motion contain the same
d'Alembertian terms of Einstein equations. this means that eq.
(\ref{eq:A3}) must satisfy the condition of describing the
characteristics of a field \emph{different} from the metric field
$g^{jk}$ -- otherwise, it becomes a void identity. --


\vskip1.80cm \small
\begin{thebibliography}{99}

\bibitem{1}
L. Pasquini \emph{et alii}, \emph{arXiv:1011.4635}
$[$astro.ph.SR$]$ 21 Nov 2010; and references therein.

\bibitem{2}
J. Cottam \emph{et alii}, \emph{Nature}, \textbf{420} (2002) 51.

\bibitem{3}
R.V. Pound and G.A. Rebka, Jr., \emph{Phys. Rev. Lett.},
\textbf{4} (1960) 337; R.V. Pound and J.L. Snider, \emph{Phys.
Rev. Lett.}, \textbf{13} (1964) 539. -- See also C. M\o ller,
\emph{The Theory of Relativity} (Clarendon Press, Oxford) 1972,
pp. 487-488.

\bibitem{4}
H. Weyl, \emph{Raum-Zeit-Materie}, Siebente Auflage
(Springer-Verlag, Berlin, \emph{etc.}) 1988, p.244; at p.322 Weyl
gives a formula for a gravitational redshift concerning arbitrary
motions of source and observer. We remark, however, the
superfluity of this result, because -- as our Author emphasizes
(see p.268) --: ``Wie die beiden K\"orper sich auch bewegen
m\"ogen, immer kann ich durch Einf\"urung eines geeigneten
Koordinatensystems die beide zusammen auf Ruhe transformieren.''

\bibitem{5}
M. v. Laue, \emph{Die Relativit\"atstheorie}, \emph{Zweiter Band}
(Friedr. Vieweg und Sohn, Braunschweig) 1956, p.111.

\bibitem{6}
E.T. Whittaker, \emph{Proc. Roy. Soc. London}, \textbf{116} (1927)
720.

\bibitem{7}
A.S. Eddington, \emph{The Mathematical Theory of Relativity},
Second Edition (Cambridge University Press, Cambridge) 1960,
p.176.

\bibitem{8}
See E.T. Whittaker and G.N. Watson, \emph{A Course of Modern
Analysis - etc.}, Fourth Edition - reprinted (The University
Press, Cambridge) 1958, Chapt. XIX.

\bibitem{9}
M. v. Laue, \emph{Phys. Zeits.}, \textbf{21} (1920) 659 -- and the
book quoted in \cite{5}, sect. \textbf{36}.

\bibitem{10}
E.T. Whittaker, \emph{Proc. Cambridge Phil. Soc.},
\textbf{24}/\textbf{\emph{I}} (1927) 32.

\bibitem{11}
T. Levi-Civita, \emph{Rend. Acc. Lincei}, s.6$^{\textrm{a}}$,
\textbf{11} (1930) 3; Idem, \emph{ibid.}, 113. -- See also: A.
Loinger, \emph{arXiv:physics/0609161} (September 19th, 2006) -- in
\emph{Appendix}: Some passages from the above memoirs by
Levi-Civita.

\bibitem{12}
A. Loinger, \emph{arXiv:physics/0606019} (June 2nd, 2006); Idem,
\emph{arXiv:0804.3991} $[$physics.gen-ph$]$ 24 Apr 2008; Idem,
\emph{arXiv:1006.3844} $[$physics.gen-ph$]$ 19 Jun 2010.

\end{thebibliography}
\end{document}